# General Quantization Rule

F. Maiz

present address:

Department of Physics, College of Science, P.O. Box9004: Saudi Arabia

King Khalid University, ABHA, Zip- 61413

permanent address:

IPEIN, B.P:62, Merazka 8000, Nabeul, Tunisia.

e-mail: fmaiz@kku.edu.sa, fethi_maiz@yahoo.fr

Abstract:

A general quantization rule for bound states of the Schrodinger equation is presented. Like fundamental theory of integral, our idea is mainly based on dividing the potential into many pieces, solving the Schrödinger equation, and deriving the general quantization rule. For both exactly and non-exactly solvable systems, the energy levels of all the bound states can be easily calculated from the general quantization rule. Using this new general quantization rule, we re-calculate the energy levels for the one-dimensional system, with an infinite square well, with the harmonic oscillator potential, with the Morse Potential, with the symmetric and asymmetric Rosen-Morse potentials, with the first Pöschl-Teller potential, with the Coulomb Potential, with the V-shape Potential, and the $ax^4$ potential, and for the three dimensions systems, with the harmonic oscillator potential, with the ordinary Coulomb potential, and for the hydrogen atom.

**I/INTRODUCTION**

After discovering Schrodinger equation, there have been many attempts for solving it. For many cases, this equation cannot be solved exactly, but it can be solved analytically for very limited number of potentials, such as Coulomb, Harmonic, oscillator and Pöschl-Teller. It may also be possible to find work carried out on quasi-exact solvable cases. Generally, researches in this area have been focused on finding some approximation methods, such as WKB.

Many quantization rules were developed, such as, Bohr-Sommerfeld [1], WKB [2-4]. Ma et all.[5] presented an exact quantization rule for one-dimensional Schrödinger equation and for three-dimensional Schrödinger equation with a spherically symmetric potential. In the exact quantization rule, in addition to $n\pi$, there is an integral term, called the quantum correction. For the exact solvable quantum systems, the energy levels can be easily solved and the solution of the ground state calculated directly from the Riccati equation [5-6]. Serrano et all.[7], propose proper quantization rule: $\int_{x_A}^{x_B} k(x)dx - \int_{x_{0A}}^{x_{0B}} k_0(x)dx = n\pi$, they carry out the exact solutions of various solvable quantum systems, and found that the energy spectra can be determined only from its ground state energy.

**II/FORMULATION**

We are going to derive the general quantization rule for the one dimensional Schrödinger equation:

$$d^2\Psi(x)/dx^2 + 2m/\hbar^2(E - V(x))\Psi(x) = 0 \qquad (1)$$

Where m is mass of the particle, ( throughout this paper, we assume that $\hbar = 1$ and $2m = 1$ ).
As described by Fig. 1,the potential energy $V(x)$ is a continuous real function (solid line) between $x = a$ and $x = b$, in order to determinate the energy levels, we start by partitioning the interval [a,b] into n+1 small subintervals $I_i = [x_i, x_{i+1}]$ each with width $h$, where $h = (b-a)/(n+1) = 2L/(n+1)$ and $x_i = a + ih$ for $i = 0,...,n$. The midpoint of each of these subintervals is given by $\rho_i = (x_i + x_{i+1})/2$. We evaluate our function $V(x)$ at the midpoint of any subinterval, and prolong it by adding two large intervals $I_{-1}$ and $I_{n+1}$ with infinite potential, so we get a bounded well, i.e.:

$$\begin{cases} \text{for } x \leq a, & x \in I_{-1} : & V(x) = \infty \\ \text{for } x_i \leq x \leq x_{i+1}, & x \in I_i : & V(x) = V(\rho_i) \\ \text{for } x \geq b, & x \in I_{n+1} : & V(x) = \infty \end{cases} \qquad (2)$$

Between $x_i$ and $x_{i+1}$, The momentum is $k_i = \sqrt{(V_i - E)}$, writing and solving equation (1) in each subinterval leads to the following solutions:

$$\Psi_i(x) = X_i \exp(k_i x) + Y_i \exp(-k_i x) \qquad (3)$$

Where $X_i$ and $Y_i$ are constants. For the two intervals $I_{-1}$ and $I_{n+1}$, the potential is infinity, and $\Psi(a) = \Psi(b) = 0$. The functions $\Psi_i(x)$ are twice derivable with respect to x, and they are continuous. The continuity of the solutions $\Psi_i(x)$ and their derivatives at different points $x_i$ allow to the elimination of $X_i$ and $Y_i$, and leads to the following real equation known as the energy quantification condition [6]:

$$B_n(E) = (c_{n,n+1} - 1)a_n P_n + (c_{n,n+1} + 1)b_n Q_n = 0 \tag{4}$$

here: $c_{ij} = c_{i,j} = \sqrt{(V_i - E)/(V_j - E)}$, $b_i = \exp(k_i h)$, $a_i = b_i^{-1}$, $Q_0 = c_{-1,0} + 1$, $P_0 = c_{-1,0} - 1$,

$P_i = (c_{i-1,i} + 1)a_{i-1}P_{i-1} + (c_{i-1,i} - 1)b_{i-1}Q_{i-1}$, and $Q_i = (c_{i-1,i} - 1)a_{i-1}P_{i-1} + (c_{i-1,i} + 1)b_{i-1}Q_{i-1}$. We note that $Q_i$ and $P_i$ are real for $V_i > E$, complex between the two turning points A and B, and $P_i = -\overline{Q_i}$.

One can proof this expression in the general case by the recurrence method as described by Maiz et all [8] and Maiz [9]. The energy levels are obtained by the energy values for which the curve of $B_n(E)$ meets the energy axis. This expression had been used to determine the superlattices band structure [7-8], and to study the Bounded Anharmonic Oscillators [10]. The energy quantification condition (eq.4) may be simplified as:

$$B_n(E) = ZT = 0 \tag{5}$$

Where $Z = Q_0 \prod_{i=0}^{n}((c_{i,i+1} - 1)\gamma_i a_i^2 + (c_{i,i+1} + 1))$, $\gamma_i = P_i/Q_i$, and $T = \prod_{i=0}^{n}(b_i)$, And Eq. (5) may be written as:

$$\operatorname{Arg}(B_n(E)) = \operatorname{Arg}(Z) + \operatorname{Arg}(T) \tag{6}$$

Here, the Arg function returns the principal value of the argument of the complex-valued expression. Since $k_i$ is real for $V_i > E$, $\arg(T) = \int_{x_A}^{x_B} \sqrt{E - V(x)}\,dx$, $x_A$ and $x_B$ are the two turning points, they satisfy the equality: $V(x_A) = V(x_B) = E_n$. Eq. (6) may be simplified to:

$$\int_{x_A}^{x_B} k(x)\,dx = \operatorname{Arg}(B_n(E)) - \operatorname{Arg}(Z) \tag{7}$$

Consider the three-dimensional Schrödinger equation with a spherically symmetric potential. After separation of the angular part of the wave function, $\psi(r) = r^{-1} R(r) Y_m^l(\theta, \varphi)$, the radial equation of the Schrödinger equation becomes:

$$d^2 R(r)/dr^2 + (2m/\hbar^2)(E - U(r))R(u) = 0,\ U(r) = V(r) + l(l+1)\hbar^2/(2mr^2) \tag{8}$$

while Eq. (8) is similar to Eq. (1), The exact quantization rule is easily generalized to the three dimensional Schrodinger equation with a spherically symmetric potential. The exact quantization rule (GQR) may be written as:

$$\int_{r_A}^{r_B} k(r)\,dr = \operatorname{Arg}(B_n(E)) - \operatorname{Arg}(Z) \tag{9}$$

In order to calculate the different terms of Eq. 7 or 9, for real (unbounded) quantum well, we use large values for $L$. The left hand side is derived manually, and the right one by using computer program.

**III/APPLICATIONS**

Using this new method, we will re-calculate energy levels of many quantum systems. The one-dimensional system with a infinite square well, where $x_A = a$, and $x_B = b$, the potential is null between the two turning points and infinity elsewhere. In this case, n=0, the solutions can be obtained easily, using Eq. (4).:

$$B_0(E) = (c_{0,1} - 1)a_0 P_0 + (c_{0,1} + 1)b_0 Q_0 = 0 \tag{10}$$

while $c_{0,1} = 0$, and $P_0 = Q_0 = I\infty$. $B_0(E) = b_0 - a_0 = 0$, we obtain $\sin(2\sqrt{E_n} L) = 0$,

the energy levels for the infinite square well are:

$$E_n = (n\pi/(2L))^2 \tag{11}$$

The potential for the one-dimensional Harmonic oscillator is $v(x) = ax^2$. It is one of the most important model systems in quantum mechanics, because an arbitrary potential can be approximated as a harmonic potential at the vicinity of a stable equilibrium point and it is one of the few quantum-mechanical systems for which an exact, analytical solution is known. The two turning points are $x_B = \sqrt{E/a}$, $x_A = -x_B$, and the two term in this new exact quantization rule (7) are:

$$\int_{x_A}^{x_B} k(x)dx = \pi E_n / \sqrt{4a} \tag{12}$$

$$\text{Arg}(Z) = \pi/2 \tag{13}$$

$$\text{Arg}(B_n(E)) = (n+1)\pi \tag{14}$$

We point out that $\text{Arg}(Z)$ is independent of the potential parameter $a$. The energy levels for the one-dimensional harmonic oscillator are:

$$E_n = (2n + 1)\sqrt{a}, \quad n=0,1,2….. \tag{15}$$

Our result agree with the previous ones.

The Morse potential is a convenient model for the potential energy of a diatomic molecule. It is a better approximation for the vibrational structure of the molecule than the quantum harmonic oscillator because it explicitly includes the effects of bond breaking, such as the existence of unbounded states.

The one dimensional Morse potential is $v(x) = D(e^{-2x/a} - 2e^{-x/a})$. The turning points are: $x_A = -a\ln(1-\sqrt{1+E_n/D})$, and $x_B = -a\ln(1+\sqrt{1+E_n/D})$.

The different term in (7) are evaluated to be:

$$\int_{x_A}^{x_B} k(x)dx = a\pi\left[\sqrt{D} - \sqrt{-E_n}\right] \tag{16}$$

$$\mathrm{Arg}(Z) = \pi/2 \tag{17}$$

$$\mathrm{Arg}(B_n(E)) = (n+1)\pi \tag{18}$$

In this case, also, we observe that $\mathrm{Arg}(Z)$ is independent of the potential parameters $a$ and $D$. Using Eq. (7), we found the energy levels for one dimensional Morse potential:

$$E_n = -(\sqrt{D} - (2n+1)/2a)^2 \tag{19}$$

The one dimensional asymmetric Rosen-Morse potential [11] is $v(x) = -U_o \mathrm{sech}^2(x/a) + U_1 \tanh(x/a)$, where $0 \leq U_1 < U_o$. This potential is useful to describing interatomic interaction of linear molecules. The two turning points are:

$$x_A = \frac{a}{2}\ln((2U_o + E - \sqrt{4U_o^2 + 4U_o E + U_1^2})/(U_1 - E)), \text{ and } x_B = \frac{a}{2}\ln((2U_o + E + \sqrt{4U_o^2 + 4U_o E + U_1^2})/(U_1 - E))$$

The different term in (7) are calculated to be

$$\int_{x_A}^{x_B} k(x)dx = \frac{a\pi}{2}\sqrt{U_o}\left[2 - \sqrt{-(E_n + U_1)/U_o} - \sqrt{-(E_n - U_1)/U_o}\right] \tag{20}$$

$$\mathrm{Arg}(Z) = \pi/2 \tag{21}$$

$$\mathrm{Arg}(B_n(E)) = (n+1)\pi \tag{22}$$

$\mathrm{Arg}(Z)$ is independent of the potential parameters $a$, $U_o$, and $U_1$, and we found the energy levels for the one dimensional asymmetric Rosen-Morse potential:

$$E_n = -(U_1/A)^2 - (A/2)^2, \text{ where } A = 2\sqrt{U_o}(1 - (2n+1)/(2a\sqrt{U_o})) \tag{23}$$

Our expression is simpler than the obtained by Ma et all.[5] and hold with $U_1 = 0$ for the symmetric Rosen-Morse potential.

The one dimensional first Pöschl-Teller potential [11] is $v(x) = \frac{1}{a^2}[\mu(\mu-1)/\sin^2(x/a) + \lambda(\lambda-1)/\cos^2(x/a)]$, where $0 < x < a\pi/2$, $\mu$ and $\lambda$ are constant greater than one, the potential tends to infinity as x tends to $0$ or $a\pi/2$. It emerges in connection with diverse physical systems, such as completely integrable many-body systems. The two turning points $x_A$ and $x_B$ satisfying $V(x_A) = V(x_B) = E_n$.

We found $x_A = a(\arctan(\sqrt{y_A}))$, and $x_B = a(\arctan(\sqrt{y_B}))$, here $y_A + y_B = (a^2 E_n - \mu(\mu-1))/(\lambda(\lambda-1)) - 1$ and $y_A y_B = \mu(\mu-1)/(\lambda(\lambda-1))$. The different term in (7) are:

$$\int_{x_A}^{x_B} k(x)dx = \frac{\pi}{2}\left[a\sqrt{E_n} - \sqrt{\mu(\mu-1)} - \sqrt{\lambda(\lambda-1)}\right] \tag{24}$$

$$\text{Arg}(Z) = \frac{\pi}{2}\left[-(\mu + \lambda - 2) + \sqrt{\mu(\mu-1)} + \sqrt{\lambda(\lambda-1)}\right] \tag{25}$$

$$\text{Arg}(B_n(E)) = (n+1)\pi \tag{26}$$

$\text{Arg}(Z)$ depends of the potential parameters $a$, $\mu$ and $\lambda$, and we found energy levels for the one dimensional first Pöschl-Teller potential:

$$E_n = ((\mu + \lambda + 2n)/a)^2 \tag{27}$$

We get the same result as Ma et all.[5].

The one dimensional Coulomb potential [12] is $v(x) = A(A-1)/x^2 - 2B/x$, for $x \geq 0$, the two turning points are $x_A = (-B + \sqrt{B^2 + A^2 E_n - AE_n})/E_n$ and $x_B = -(B + \sqrt{B^2 + A^2 E_n - AE_n})/(E_n)$, the different term in (7) are evaluated to be:

$$\int_{x_A}^{x_B} k(x)dx = \pi\left[IB/\sqrt{E_{nl}} + \sqrt{A(A-1)}\right] \tag{28}$$

$$\text{Arg}(Z) = -\pi\left[(A-1) + \sqrt{A(A-1)}\right] \tag{29}$$

$$\text{Arg}(B_n(E)) = (n+1)\pi \tag{30}$$

$\text{Arg}(Z)$ depends on the potential parameter $A$, and we found the energy levels for the one-dimensional Coulomb potential:

$$E_n = -B^2/(A+n)^2 \tag{31}$$

Our result fully agree with those obtained by Ho[12].

The one dimensional V-shape potential is $V(x) = a|x|$, where $a \geq 0$, the two turning points are:

$x_A = -E_n/a$ and $x_B = E_n/a$, in this case:

$$\int_{x_A}^{x_B} k(x)dx = 4E_n^{3/2}/(3a) \tag{32}$$

$$\text{Arg}(Z) = \pi/2(1 + 130\sin(\pi(n + 0.9997))e^{-0.85n}) \tag{33}$$

$$\text{Arg}(B_n(E)) = (n+1)\pi \tag{34}$$

$\text{Arg}(Z)$ depends only on the quantum number $n$ and we found the energy levels as:

$$E_{nl} = (3\pi a/8 (2n + 1 - 130\sin(\pi(n + 0.9997))e^{-0.85n}))^{2/3} \tag{35}$$

For large values of n, we obtain the same result as the WKB quantization rule such as: $E_{nl} = (3\pi a / 8(2n+1))^{2/3}$.

The one dimensional $ax^4$ potential is $V(x) = ax^4$, where $a \geq 0$, the two turning points are:

$x_A = -(E_n/a)^{1/4}$ and $x_B = (E_n/a)^{1/4}$, in this case:

$$\int_{x_A}^{x_B} k(x)dx = 1.748 E_n^{3/4} / a^{1/4} \tag{36}$$

$$\mathrm{Arg}(Z) = \pi/2(1 - 0.038/(n+0.23)) \tag{37}$$

$$\mathrm{Arg}(B_n(E)) = (n+1)\pi \tag{38}$$

$\mathrm{Arg}(Z)$ depends only on the quantum number n and we found the energy levels as:

$$E_n = (9/10\ (2n+1 - 0.038/(n+0.23)))^{4/3} a^{1/3}. \tag{39}$$

The effective potential for the three-dimensional harmonic oscillator is $V_{\mathrm{eff},l} = l(l+1)/r^2 + ar^2$, the two turning points are $r_A = \sqrt{(E_{nl} - \sqrt{E_{nl}^2 - 4al^2 - 4al})/(2a)}$ and $r_B = \sqrt{(E_{nl} + \sqrt{E_{nl}^2 - 4al^2 - 4al})/(2a)}$,

the different term in (9) are calculated to be:

$$\int_{x_A}^{x_B} k(x)dx = \pi/2 \left[ E_{nl}/(2\sqrt{a}) - \sqrt{l(l+1)} \right] \tag{40}$$

$$\mathrm{Arg}(Z) = \frac{\pi}{2}\left[\sqrt{l(l+1)} + 1/2 - 1\right] \tag{41}$$

$$\mathrm{Arg}(B_n(E)) = (n - l + 2)\pi/2 \tag{42}$$

$\mathrm{Arg}(Z)$ depends on the potential parameter l, and we found the energy levels for three-dimensional harmonic oscillator:

$$E_{nl} = (2n+3)\sqrt{a} \tag{43}$$

The ordinary Coulomb potential is $V_l(r) = l(l+1)/r^2 - e^2/r$, where e the electric charge, and l the orbital angular quantum number. This potential expression is similar to the one-dimensional Coulomb one, so, let us take $A = l+1$, $2B = e^2$, and profit of the obtained result $E_n = -B^2/(A+n)^2$. We get exactly the same result as Ho[12]:

$$E_{nl} = -e^4/(4(n+l+1)^2) \tag{44}$$

The effective potential for hydrogen atom is $V_l(r) = l(l+1)\hbar^2/(2mr^2) - e^2/r$, where m the electron mass. This potential expression is, also similar to the one-dimensional Coulomb one, so, let us take $A = l+1$, $2B = 2me^2/\hbar^2$, and find the energy levels $E'_{nl} = 2mE_{nl}/\hbar^2$. we get $E'_{nl} = -(2me^2/\hbar^2)^2/(n+l+1)^2$, and we found energy levels for the effective potential for hydrogen atom:

$$E_{nl} = -(2m/\hbar^2)e^4/(n+l+1)^2 \qquad (45)$$

If $n = l+1$ We obtain exactly the same result as Ma et all.[6].:

$$E_{nl} = -me^4/(2n^2\hbar^2) \qquad (46)$$

**III/CONCLUSION**

In this paper, we present a general quantization rule for the bound states of one-dimensional Schrödinger equation and three-dimensional Schrödinger equation with a spherically symmetric potential. Our idea is mainly based on dividing the potential into many pieces, solving the Schrödinger equation, and deriving the exact quantization rule. For exactly solvable systems, the energy levels of the quantum systems can be easily derived from the general quantization rule. The results are found in agreement with the previous obtained ones. For the non exactly savable systems, the general quantization rule gives numerically the exact energy levels.

The general quantization rule is constituted by three terms. First one is the integral of the momentum between the two turning points, it's calculated manually. The two second ones are, $\text{Arg}(B_n(E))$, which depends on quantum numbers, and $\text{Arg}(Z)$, it's generally function of the potential parameters and the quantum numbers in some cases, are derived by using computer programs. This general quantization rule is exact, because it solves with high degree of accuracy all energy levels of all the bound states for both exactly and non exactly solvable systems, and like the Fundamental theory of integral based on the same idea.

Figure caption: Fig. 1: potential curve, the two turning points are A and B.

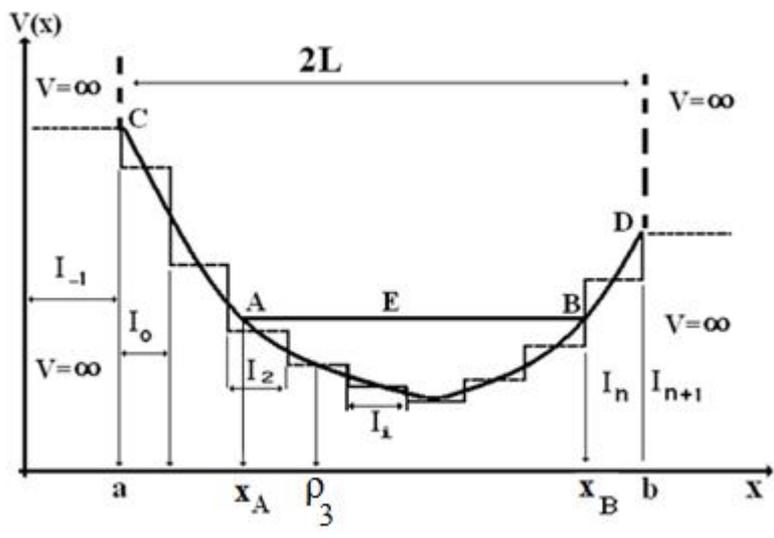